%% file: cloudpg.tex
\documentclass{IEEEtran}

\usepackage[utf8]{inputenc}
\usepackage[T1]{fontenc}
\usepackage{cite}
\usepackage{amsmath,amssymb,amsfonts}
\usepackage{algorithmic}
\usepackage{graphicx}
\usepackage{textcomp}
\usepackage{enumitem}
\usepackage{xcolor}
\usepackage{hyperref}
\usepackage{listings}
\usepackage{tikz}
\usepackage{textcomp}
\def\BibTeX{{\rm B\kern-.05em{\sc i\kern-.025em b}\kern-.08em
    T\kern-.1667em\lower.7ex\hbox{E}\kern-.125emX}}

\begin{document}

\title{Cloud Property Graph: Connecting Cloud Security Assessments with Static Code Analysis \thanks{\textcopyright 2021
		IEEE. Personal use of this material is permitted. Permission from IEEE must be obtained for all other uses, in any
		current or future media, including reprinting/republishing this material for advertising or promotional purposes,
		creating new collective works, for resale or redistribution to servers or lists, or reuse of any copyrighted
		component of this work in other works. DOI:
		\href{https://dx.doi.org/10.1109/CLOUD53861.2021.00014}{10.1109/CLOUD53861.2021.00014}. Funded by the Horizon 2020
		project MEDINA, grant agreement ID 952633. } }

\author{\IEEEauthorblockN{Christian Banse, Immanuel Kunz, Angelika Schneider and Konrad Weiss}\\
	\IEEEauthorblockA{
		Fraunhofer AISEC, Garching b. München, Germany\\
		Email: \{firstname.lastname\}@aisec.fraunhofer.de}
}

\maketitle

\begin{abstract}
	In this paper, we present the Cloud Property Graph (CloudPG), which bridges the gap between static code analysis and
	runtime security assessment of cloud services. The CloudPG is able to resolve data flows between cloud applications
	deployed on different resources, and contextualizes the graph with runtime information, such as encryption settings.
	To provide a vendor- and technology-independent representation of a cloud service's security posture, the graph is
	based on an ontology of cloud resources, their functionalities and security features.  We show, using an example,
	that our CloudPG framework can be used by security experts to identify weaknesses in their cloud deployments,
	spanning multiple vendors or technologies, such as AWS, Azure and Kubernetes. This includes misconfigurations, such
	as publicly accessible storages or undesired data flows within a cloud service, as restricted by regulations such as
	GDPR.
\end{abstract}


\IEEEpeerreviewmaketitle

\lstset{basicstyle=\footnotesize}

\section{Introduction}
\label{introduction}
\input{sections/introduction.tex}



\section{Systemizing Cloud Resource Properties}
\label{ontology}
\input{sections/ontology.tex}

\section{Framework Architecture}
\label{approach}
\input{sections/approach.tex}
\section{Prototype Implementation}
\label{implementation}
\input{sections/implementation.tex}

\section{Evaluation and Discussion}
\label{sec:discussion}
\input{sections/discussion.tex}

\section{Related Work}
\label{relatedwork}
\input{sections/related.tex}

\section{Conclusions}
\label{conclusions}
\input{sections/conclusions.tex}

\bibliographystyle{unsrt}
\bibliography{cloudpg.bib}

\end{document}

%% file: sections/introduction.tex

\newenvironment{requirement}{\begin{enumerate}[label=\textbf{D\arabic*}]\raggedright}{\end{enumerate}}

Analyzing the security of a cloud service, from the virtual infrastructure it is deployed on, up to the application code
that implements the actual service, is a complex task involving multiple challenges. First, there is an ever-growing
variety of virtual infrastructure services and cloud vendors in the cloud ecosystem. Each cloud resource has a unique
set of properties, that needs to be checked, e.g. to find weaknesses or to prepare for a cloud service certification.
Especially when dealing with multi-cloud or hybrid-cloud scenarios, selecting the correct security properties is a
tedious task. All major public cloud vendors offer extensive APIs to retrieve such configurations and logging
information. However, the semantics and naming of relevant properties, such as encryption or access control
configuration, are inconsistent across different cloud vendors.

Second, next to the configured cloud resources, the cloud service consists of the actual application code. While static
application security testing (SAST) tools can be used to assess it, significant challenges remain in analyzing
applications that are deployed as part of a cloud service. For example, developers of a function within the service
might make certain assumptions at design time about the runtime environment with regards to encryption or
authentication. If the surrounding infrastructure changes---as it frequently does in a cloud environment---these
assumptions might not hold at runtime and lead to weaknesses in the overall system. A shortcoming of static code
analysis is therefore, that no---or insufficient---information about its runtime environment is available during
analysis.

This challenge also extends to the analysis of data flows across different components or services. An undesired data flow might
occur to an application that is deployed in a specific region which, however, can only be determined
by including deployment properties in the analysis. Additionally, the overall program logic may be fragmented across components. 
In the case of serverless functions only a small subset of the actual executed code (the function itself) is visible to
an analysis tool. The majority of the behavior, such as triggers or data sinks that the function interacts with, remains
hidden to such tools. Therefore, they may not be able to identify data flows that can lead to the compromise of a
service, for example through an improper invocation of a serverless function.

To address these issues, we present the \textit{Cloud Property Graph} (CloudPG). It is an extension of a Code Property
Graph (CPG)~\cite{yamaguchi2014modeling}, which itself is a labeled directed graph, representing source code. A CPG
generates a language-independent representation of an application's structural components, i.e. classes or methods, as
well as information about data flows or program dependence. To build the CloudPG, we enrich this
graph with additional nodes and edges that represent the actual deployment of the code as service(s) in the cloud at
runtime.


In doing so we aim to adhere to the following design goals: Our comprehensive graph should allow for an in-depth
analysis of the deployed service and enable building a service-independent representation of a cloud deployment
(\textbf{DG1}). It should bridge the gap between static code analysis and runtime assessment (\textbf{DG2}), while
allowing for tracking data flows across different micro-services including interactions between cloud resources
(\textbf{DG3}). Lastly, by providing an efficiently searchable representation of the results can be used to verify
requirements, properties and relationships of components, for instance a proper encryption configuration (\textbf{DG4}).



In summary, in the course of the paper we present the following contributions:
\begin{itemize}
	\item an ontology that represents cloud services, cloud-related software frameworks, their resources and security
	      properties, as well as instantiations for AWS, Microsoft Azure, Kubernetes and popular Web-based
	      libraries,
	\item an analysis framework, which combines aspects of cloud workload security and static code analysis, which
	      allows to query security-related properties of cloud-based services, independently of the underlying
	      provider, and
	\item a prototype implementation of the proposed framework.
\end{itemize}


%% file: sections/ontology.tex


In this section, we propose an abstract representation of cloud resources and their security properties. 
We provide this abstraction (design goal \textbf{DG1}) in the form of an
ontology\footnote{Published at \url{https://github.com/clouditor/cloud-property-graph}}.

\subsection{An Ontology for Cloud Resources}



Our ontology consists of three separate taxonomies for \textit{cloud resources}, \textit{cloud-related software
frameworks}, as well as their \textit{functionalities and security features}. It further establishes relationships
between them, describing which cloud resources offer which security features.

\subsubsection{Cloud Resource Taxonomy}
\begin{figure}[t]
	\centering
	\includegraphics[width=0.89\linewidth, keepaspectratio]{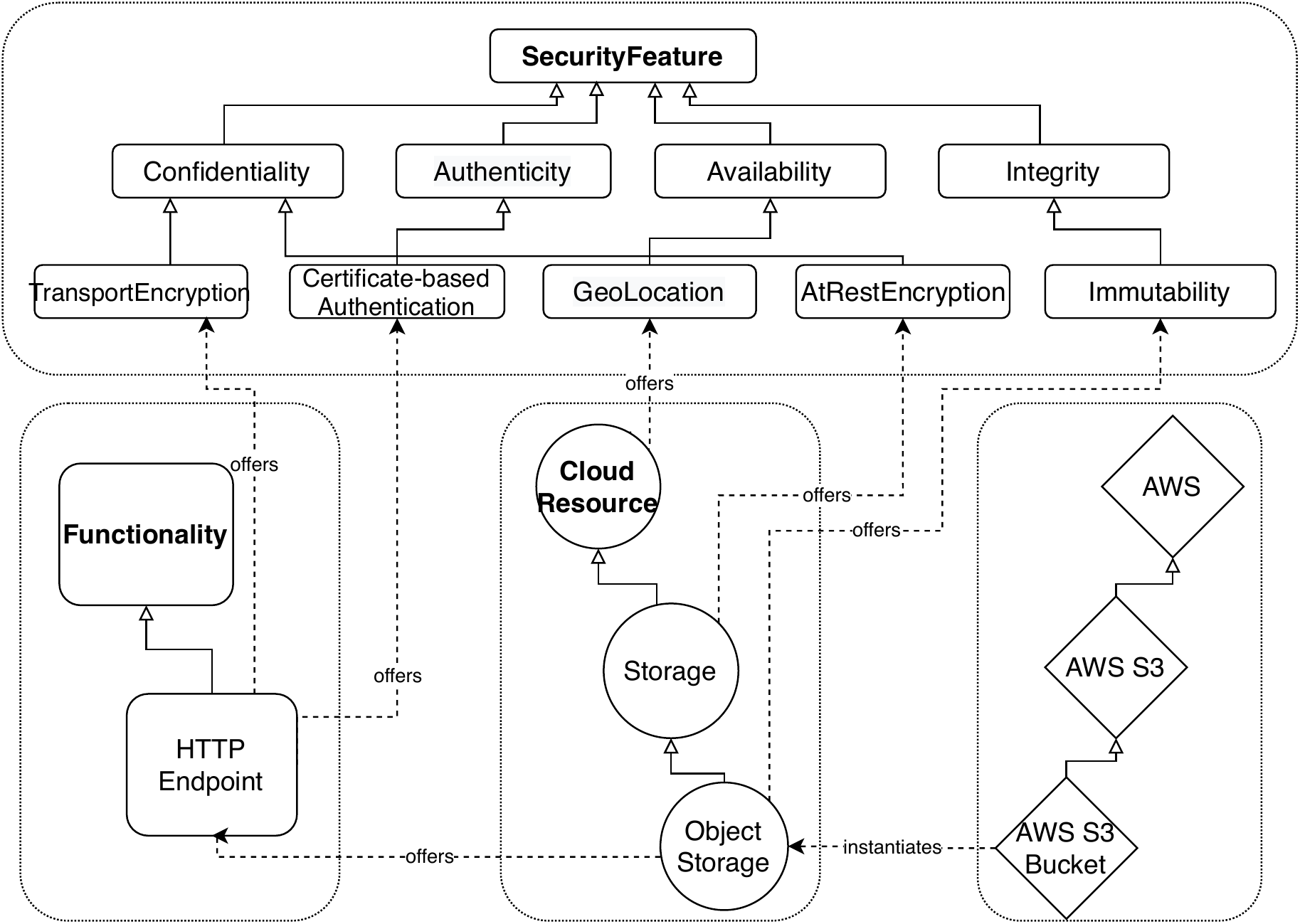}
	\caption{An excerpt of the ontology and its instantiation: An AWS S3 Bucket is an instantiation of the abstract resource type ObjectStorage which in turn is a child node of Storage. As Storage offers the security feature AtRestEncryption, it can be reasoned that an AWS S3 Bucket must also offer this feature.}
	\label{fig:ontology}
\end{figure}

For the cloud resources themselves we first consider traditional cloud offerings, such as compute, storage, networking,
identity management and logging. An example inheritance can be seen in Figure \ref{fig:ontology} where an
\textit{ObjectStorage} has a generic \textit{Storage} as its parent which in turn has the \textit{CloudResource} entity
as its parent. We also model cloud-related CI/CD resources, such as jobs and workflows.

\subsubsection{Software Frameworks}
Next, we identify a taxonomy that evolves around the usage of different software frameworks. This includes libraries
typically used in a cloud or distributed context, e.g. web service frameworks, HTTP client libraries, loggers,
authentication handlers or the Cloud SDKs to access cloud resources themselves.

\subsubsection{Security Features and Functionalities Taxonomy}
Lastly, with this taxonomy we aim at extracting functionalities and security features that are common across cloud
resources and software frameworks. Functionalities are more generic in nature and are used to describe certain
characteristics about a service, e.g. that an object storage typically has an HTTP endpoint to access it. Application
code can have functionalities as well: for instance an HTTP client library will have the functionality to issue HTTP
requests. 


We collect security features that are offered by different cloud vendors and assign them to commonly used security
properties, based on the STRIDE methodology\cite{howard2006security} as follows.

\textit{Confidentiality}, e.g. at-rest encryption and transport encryption options; \textit{integrity}, e.g.
immutability of storage resources; \textit{availability}, e.g. backups or geographic location; \textit{authentication},
e.g. password-based authentication; \textit{authorization}, e.g. access restrictions of IP addresses and ports; and
\textit{auditing}, e.g. audit log output. 

While it is impossible to say if this taxonomy is complete, all protection goals have at least one feature assigned.
Specific data properties, e.g. the used encryption algorithm, further detail those security features. We model the
relationship between cloud resources or frameworks and the features they offer using the object property \textit{offers}
(see Figure \ref{fig:ontology}). 

\subsection{Instantiating the Ontology}
\label{sec:instance}
By separating the abstract ontology from concrete instantiations, we create a modular structure of a long-lived
abstraction on the one hand and an adaptable and maintainable instantiation on the other. We have created instantiations
of the proposed cloud resource ontology for AWS and Azure, as well as for Kubernetes resources.
For example an AWS EC2 Volume is an instantiation of the class \textit{BlockStorage}. 
While the instantiation of most AWS and Azure resources is straightforward, since
both follow similar concepts in their service offering, the instantiation of Kubernetes resources is more complex. 

We model a Kubernetes Container as the central computing resource (similar to a function or a virtual machine) and map
some special Kubernetes concepts, like an Ingress resource 
onto the cloud resources 
that are
functionally equal. For instance, an Ingress is an instantiation of a \textit{LoadBalancer} since it provides the
capability of receiving traffic and distributing it to the respective containers.
While this instantiation may not be perfect in its functional details, we do achieve
our goal of enabling the graph-based representation of these resources and their security features.


Furthermore, we instantiate the software framework's taxonomy using popular libraries from the Java, Python and
JavaScript ecosystem. For example, we classify the software library Jersey as a \textit{HttpClientLibrary}, used to
execute GET and POST requests (\textit{HttpRequest} functionality in the ontology) to web services. We connect the HTTP
request to a \textit{CallExpression} in the CPG, representing a function call in code. On the server-side we classify
that Spring/SpringBoot can be used as an \textit{HttpServer} framework, which offers several HTTP-related
functionalities. Using Spring, Java classes can be annotated with the \verb|RestController| annotation and are thus used
as a \textit{HttpRequestHandler}, with each method in the class (further annotated with \texttt{RequestMapping}) usually
serving as an \textit{HttpEndpoint}, representing a certain URL, for example \texttt{/hello}.

Lastly, we include examples in our ontology instantiation with regards to Cloud SDKs, specifically to the Azure Storage
SDK. We model accesses to an Azure storage account by instantiating the \textit{ObjectStorageRequest} class, holding
references to an \textit{ObjectStorage} and other data properties, such as the access type (read, write, create).



%% file: sections/approach.tex


\begin{figure}
	\centering
	\includegraphics[width=1\linewidth, keepaspectratio]{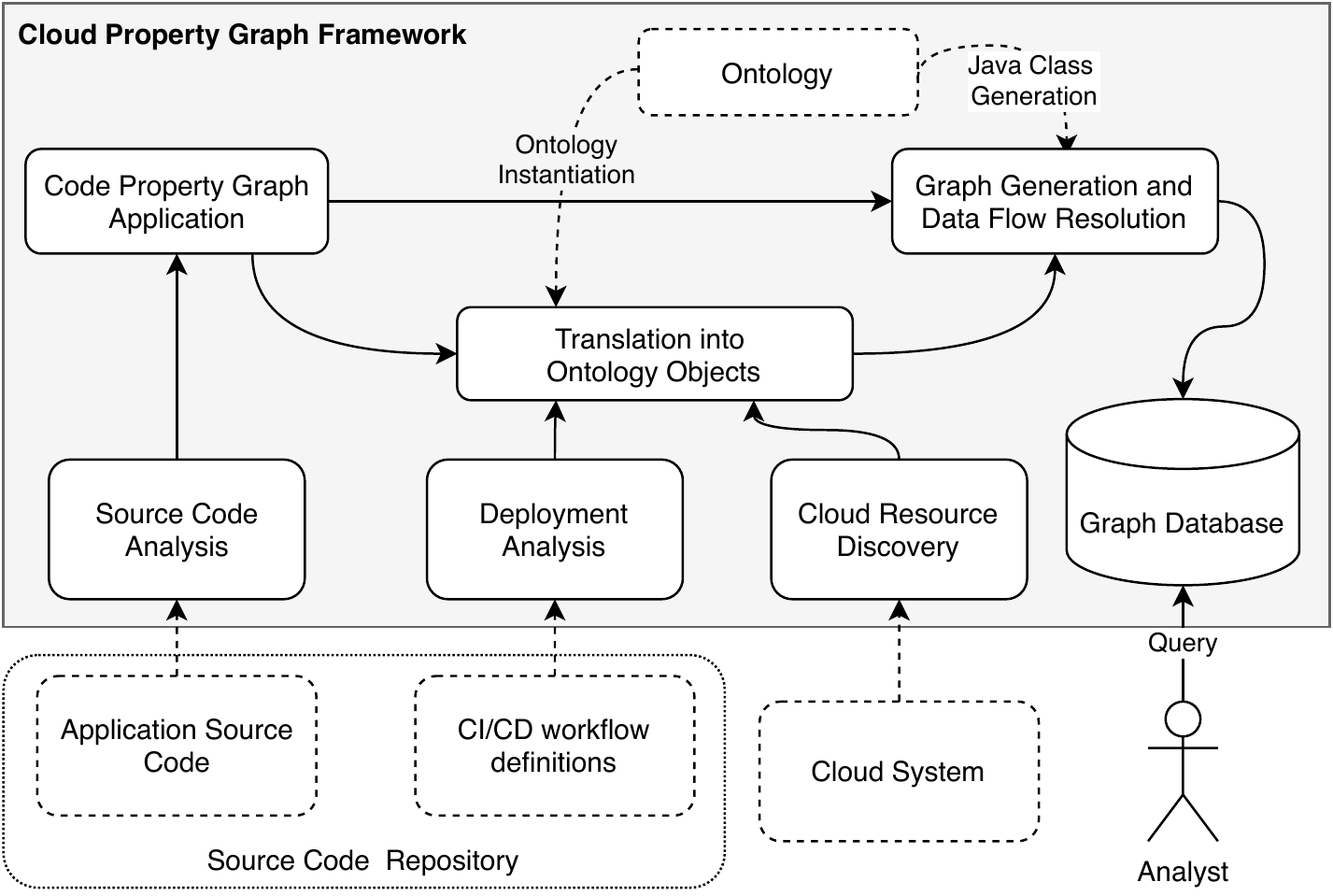}
	\caption{Architecture of the proposed CloudPG Analysis Framework}
	\label{fig:architecture}
\end{figure}

The primary goal of our framework is to assess security properties of software applications
deployed on cloud systems, based on our ontology (\textbf{DG1}). It bridges the gap between static
code analysis and runtime assessment (\textbf{DG2}) and therefore merges different data
sources, as described below:
\begin{itemize}
	\item Static code analysis of the actual application code,
	\item an analysis of CI/CD workflow definitions that deploy said application to a target cloud system, and
	\item information about cloud workload security, i.e. the security of the provisioned cloud resources in the deployment.
\end{itemize}

Furthermore, the framework contains specific analysis modules dedicated to the resolution of data flows (\textbf{DG3}) 
across the cloud service. Lastly, it persists the generated CloudPG into a searchable graph database (\textbf{DG4}).

\subsection{Bootstrapping the Cloud Property Graph}

Figure~\ref{fig:architecture} shows the architecture of the framework, including its modules. First, a list of possible
source code repositories which are deployed as part of the cloud service needs to be identified. The source code is
then translated into the language-independent representation of a CPG following the regular process of
graph-based static code analysis.


We extend an existing CPG by importing OWL-based ontologies to build up new node and edge classes in the graph. This
forms the basis for our CloudPG. Additionally, in the first step, relevant cloud frameworks and their functionalities
are identified in the source code. This includes frameworks, such as REST controllers or libraries for web-based
requests. 
The existence of such frameworks is modelled as objects represented in the ontology and thus in
the graph, e.g. using a \textit{HttpRequest} node. Furthermore, specific properties, such as individual endpoints
of a REST controller in the application are modelled as well, e.g. using an \textit{HttpEndpoint} node. This is later
used as a link to other network services, such as load balancers and cluster ingress endpoints, to connect the data flow
from a deployed URL to the particular method in an application code that handles it.

\subsection{Deployment Analysis and Discovery of Cloud Resources}
\label{sec:discovery}

In addition, information about the deployment of the target application(s) is gathered using specific APIs offered by
the various DevOps platforms well as deployment files. They are analyzed with regards to references to cloud
deployments, such as container images, container orchestration systems, such as Kubernetes, or cloud resource
deployments, like virtual machines. All these serve as indications that the analyzed application is deployed into the
identified cloud system. Note that they are modelled as terms in the ontology as well. 

\begin{figure*}[h!]
	\centering
	\includegraphics[width=0.95\linewidth, keepaspectratio]{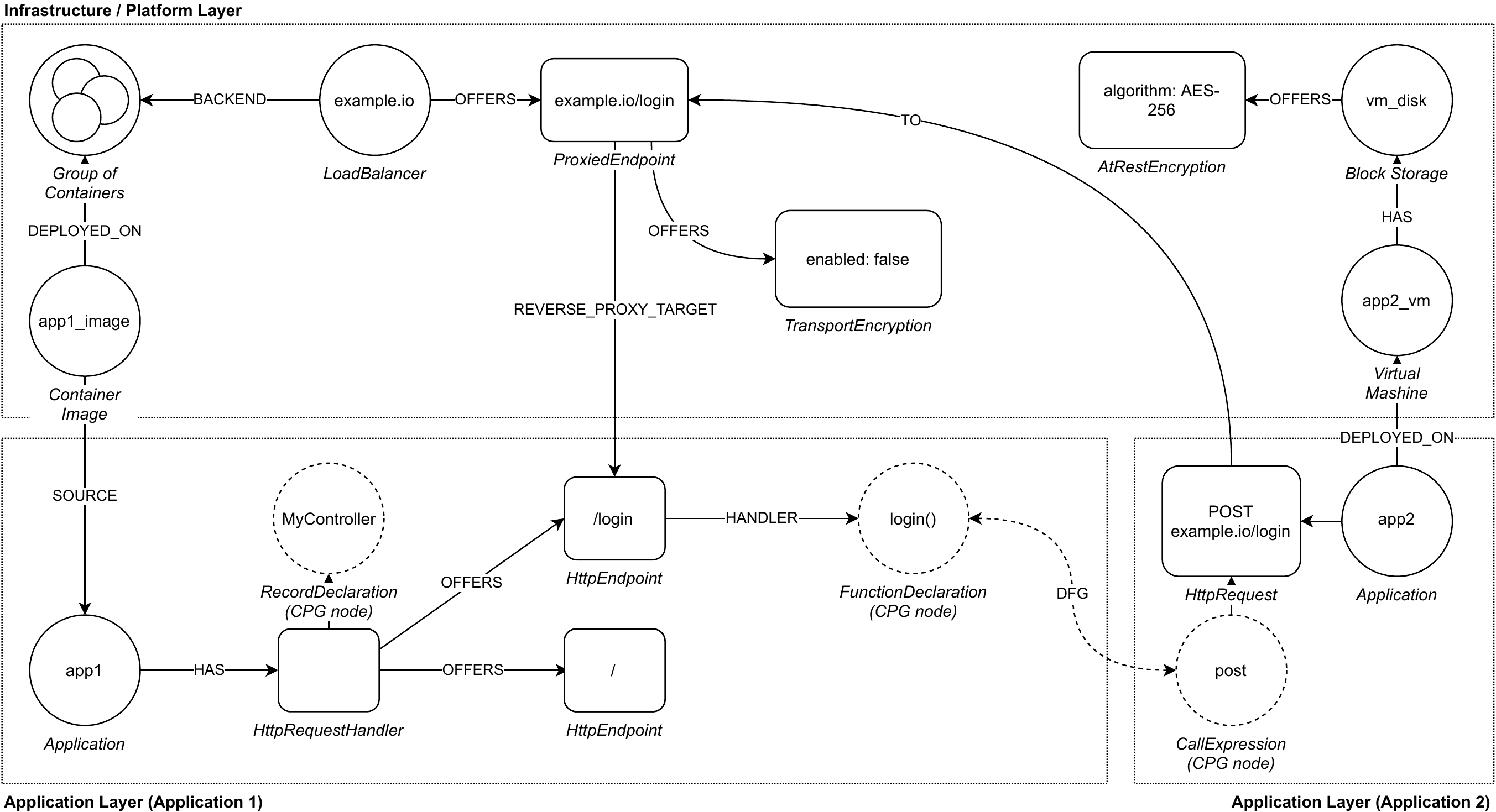}
	\caption{An excerpt from the CloudPG showing two applications \textit{app1} and \textit{app2}: app1 include a web
		 service offering two endpoints, \textit{/login} and \textit{/}. The endpoints are exposed through a load
		 balancer, serving the URL \textit{example.io}. Therefore, an additional \textit{ProxiedEndpoint} node is
		 contained in the graph, representing the deployed web service, e.g., on \textit{example.io/login}. app2 is
		 deployed on a VM and contains an http request to that particular web resource. Thus, the data flow can be
		 connected between the \textit{CallExpression} of the HTTP POST call to the \textit{FunctionDeclaration}. The
		 graph also shows an additional security feature \textit{TransportEncryption} connected to the HTTP endpoints.}
	\label{fig:cloudpg-graph-dataflow}
\end{figure*}

In the next step, all relevant cloud resources and their properties in the deployment target system(s)
are discovered and modelled as terms on the instantiated.
First, for a particular cloud resource, the cloud provider-specific type is determined. In this example, we assume a
cloud resource named \textit{myvolume} of type \textit{AWS EC2 Volume}. Then we look up the type in the instantiated
ontology for AWS (see Section~\ref{sec:instance}). Following the example, we discover that this resource is
classified as a \textit{BlockStorage} and thus create an appropriate node in the graph representing this class. In the
next step, basic data properties of the node, such as its name, are populated. Second, specific security features are modelled for the particular resource. By looking into the instantiated ontology,
we learn that block storages offer \textit{AtRestEncryption} and possess a \textit{GeoLocation} attribute. We use that
to create an appropriate node in the graph as well and connect it to our \textit{BlockStorage} node. Finally, further
properties of the security features are populated, like the correct geographic region or a configured encryption algorithm.
See Figure~\ref{fig:cloudpg-graph-dataflow} for an example graph.

\subsection{Resolution of Data Flows}
\label{sec:architecture-data-flows}

To resolve data flows (see \textbf{DG3}) of the cloud service, we differentiate between several typical scenarios.

\paragraph{Direct HTTP requests from an application to a web service, e.g. within the same cluster or virtual network}

In general, this includes data flows to cloud resources offering an HTTP backend, i.e. a publicly accessible object
storage. To resolve this data flow, we connect \textit{HttpRequest} nodes in the graph to nodes that offer the ontology
functionality \textit{HttpEndpoint}, matching their URL, paths and the HTTP method.

\paragraph{HTTP requests via a load balancer} Load balancers use a reverse proxy to connect to the backing web service.
In our ontology, we model this as a \textit{ProxiedEndpoint}, which inherits from a regular \textit{HttpEndpoint}. We
traverse the graph, identifying all \textit{HttpEndpoint}s that belong to an application which are the target of a load
balancer. We then create \textit{ProxiedEndpoint} for each identified ``local'' HTTP endpoint and prefix the URL of the
load balancer. Afterwards, the normal resolution method described above can be used.

\paragraph{Requests to Cloud resources within an application}

We identify relevant expressions in the CPG that represent operations of a \textit{CloudSDK} and resolve identifiers of
cloud resources in the source code to nodes in the graph. For example, writing to an object storage, such as AWS S3,
would result in an \textit{ObjectStorageRequest} node connected to an \textit{ObjectStorage} node.

%% file: sections/implementation.tex


In this section, we present a Kotlin-based implementation of the proposed architecture, built as an open-source
project\footnote{\url{https://github.com/clouditor/cloud-property-graph}}. It consists of a complete analysis framework
including the functionality to persist and query the CloudPG using the Neo4J graph database. It leverages the
\textit{cpg}\footnote{\url{https://github.com/Fraunhofer-AISEC/cpg}} project to bootstrap a code property graph, which
is then enriched by further edge and node types to build the CloudPG. An analyst can then later use the query language
Cypher to query the persisted graph. Section~\ref{sec:example} lists several example queries.

\subsection{CloudPG Modules and Supported Languages}

We use the information identified in the ontology in multiple steps of the analysis platform. Since the \textit{cpg}
library is Java-based, additional node (and edge) types in the graph are represented in Java classes, which all inherit
from the same \texttt{Node} base class. We therefore automatically generate appropriate class files based on the
ontology modeled in Section~\ref{ontology}. This allows us to focus on modelling the properties and structural
relationships in the ontology rather than implicitly modelling a semantic model in a programming language. It also helps
the ontology to be implementation-agnostic.
%
%
%

%
In addition to the languages already supported by the original CPG implementation (Java, C/C++, Python and Go), we added
basic support for Ruby and JavaScript/NodeJS.
The modules in the architecture are implemented as \textit{passes}, which are automatically called during the graph
construction. 

\subsection{Discovering CI/CD Workflows}
One such pass analyzes CI/CD workflows for images that are created and pushed to container registries. This information
is then used to connect an application to a compute resource. In the concrete implementation we scan workflow definition
files from GitHub Actions. They are YAML-based and follow a specific syntax enabling the execution of jobs and steps
within a job\footnote{\url{https://docs.github.com/en/actions}}. In particular, we are interested if a workflow step is
building a Docker image using the \texttt{docker} CLI command. If so, we create an \textit{ContainerImage} node in the
graph and populate it with the appropriate properties.
\subsection{Cloud-specific Discovery Passes}
%
%
We implemented the Cloud discovery and ontology mapping for a limited set of cloud resources within Microsoft Azure and
AWS. In particular, we focus on the enumeration of containers, container clusters, virtual machines, storage accounts
and volumes, including their associated security property. Specifically, we use the cloud providers' SDKs to retrieve
the individual Cloud resources as Java/Kotlin objects represented in SDK-specific classes (such as
\texttt{com.azure.(...).models.Disk}). We then create matching nodes in the CloudPG based on the type defined in the
instantiated Ontology for the particular provider (see Section~\ref{sec:instance}). 
%
%
Furthermore, a similar pass exists that uses the Kubernetes SDK to retrieve pods, services and ingress definitions
from a Kubernetes cluster to create nodes that represent compute resources, for example a \textit{Container} or a \textit{LoadBalancer}. 
The \textit{Container} nodes are also connected to \textit{Image} nodes, which may already exist from previous passes,
serving as a link to connect the container to an application through its image.
%
%
\subsection{Data Flow Resolution}
%

We implemented modules for CloudPG data flow resolution of several popular Web frameworks, namely Spring and JAX-RS for
Java, Flask for Python, WEBrick and HttpDispatcher for JavaScript/NodeJS. In the following, we use Spring to demonstrate
the approach, which is fairly similar for the aforementioned frameworks.

\paragraph{Preparing Data Flow Resolution}

We start preparing the (HTTP) data flow resolution by building up \textit{HttpEndpoint}s and their associated
\textit{HttpRequestHandler}s. According to the instantiated ontology for Spring (see Section~\ref{sec:instance}), the
library can be used to launch an \textit{HttpServer}, which handles individual HTTP requests through a controller class.
This Java class, annotated with \texttt{@RestController}, is represented as a \textit{RecordDeclaration} in the original
CPG and as an \textit{HttpRequestHandler} node in the CloudPG. Futhermore, individual methods in this class, annotated
by \texttt{@RequestMapping}, are modelled as an \textit{HttpEndpoint} and handle HTTP requests belonging to a certain
URL, e.g. \texttt{/login}. This can be seen in Figure~\ref{fig:cloudpg-graph-dataflow} on the bottom left side.

If the application is connected to a load balancer, we create further \textit{ProxiedEndpoint} nodes that represent the
endpoints in the load balancer, as described in Section~\ref{sec:architecture-data-flows}. This connection can be seen
in the middle of Figure~\ref{fig:cloudpg-graph-dataflow} in the nodes \textit{/login} (HttpEndpoint) and
\textit{example.io/login} (ProxiedEndpoint), connected to the load balancer of \textit{example.io}.
%
%
%
\paragraph{Resolving HTTP-related Data Flows}

After the preparation phase, data flows originating from a data source (usually an HTTP request) to a known data sink,
such as a REST API, can be connected. Use cases include invoking remote function calls from one micro-service to
another, e.g. for authentication. In particular, the reference implementation can analyze such calls for the Jersey
HTTP library for Java as well as the Requests library for Python. 
%
%
In a next step, independently of the underlying technical implementation, all \textit{HttpRequest} nodes are connected
to suitable \textit{HttpRequest}s or \textit{ProxiedEndpoint}s, i.e. those that match the URL and HTTP method.
Furthermore, DFG edges are added between the function declarations that handle the endpoint and the call expression
representing the HTTP request.

%
\paragraph{Resolving Cloud Storage-related Data Flows}
Lastly, we use the Azure Storage SDK as an example demonstrating the resolution of data operations made by a cloud SDK
within an application to its specific cloud resources. Similarly, to the previous step we identify Java objects
representing a client, configured with a URL of the storage account in a builder-pattern style. This client is then in
turn used to issue operations, such as \texttt{create()} or \texttt{append()} files in the storage account container.
The individual operations are modelled as \textit{ObjectStorageRequest} and connected to the appropriate
\textit{ObjectStorage}, identified in the creation of the client using its URL endpoint and name, as well as to the call
expression in the CPG.
%

%% file: sections/discussion.tex


In this section, we present several generalizable example weaknesses that can be identified in a deployed testbed with
our framework and discuss them in comparison to alternative solutions. We also present basic performance measurements.

\begin{table}
  \caption{Weaknesses to be evaluated in the Testbed}
  \centering
  \label{tab:weaknesses}
  \begin{tabular}{lp{5.9cm}}
    \hline
    \textbf{ID} & \textbf{Description}\\
    \hline
    $DataExposure_1$ & The \textit{am-containerlog} is misconfigured to have public access enabled.\\
    \hline
    $Confidentiality_1$ & The \textit{am-containerlog} resource is using TLS 1.1.\\
    \hline
    $DataFlow_1$ & The VM \textit{ratings} micro-service is located in an AWS US region; whereas the rest of the
    services are in Europe and data should stay in Europe.\\
    \hline
    $DataFlow_2$ & The GitHub Actions pushes Docker images to the GitHub Container Registry, which is assumed to be in
    the US. AKS, located in Europe, retrieves them.\\
    \hline
    $DangerousLog_1$ & The \textit{productpage} service was adjusted to log the contents of the \textit{login} request
    (username, password).\\
    \hline
    $DataExposure_2$ & The AKS cluster is configured to forward container logs to an Azure Log Analytics Workspace.
    These log files are persisted onto a Azure Storage Account Container (\textit{am-containerlog}). Because of
    $DataExposure_1$, all container logs are public.\\
    \hline
    $DataExposure_3$ & Through the combination of $DangerousLog_1$ and $DataExposure_2$, passwords are logged onto a
    publicly accessible storage container.\\
    \hline
  \end{tabular}
\end{table}



For our testbed, we employed the
\textit{bookinfo}\footnote{\url{https://github.com/istio/istio/tree/master/samples/bookinfo}} example from the Istio
framework, which is a Cloud-based service, divided into four micro-services. We distributed its \textit{productpage},
\textit{details} and \textit{reviews} micro-services to an Azure Kubernetes cluster and the \textit{ratings}
micro-service to an AWS EC2 instance. We also added an automated deployment using GitHub Actions.
The source code has a total of 3864 LoC spread across four languages (Python, Ruby, Java and JavaScript/NodeJS). This
translates into about 1800 nodes in the CloudPG. The cloud resources are provisioned with a variety of different
configurations, deliberately leading to several weaknesses, described in Table~\ref{tab:weaknesses}.

We executed the analysis of the \textit{bookinfo} service 10 times and recorded the times using benchmarking tools built
into the cpg library. The analysis was performed on an Azure virtual machine (\text{b2s flavor}) using 2 vCPUs and 4 GB
RAM. The overall CloudPG construction took \textit{19,2s}, with the majority of time spent in resource discovery of
Azure and AWS resources (\textit{10,5s}) as well as the translation of source code (\textit{5,3s}). Persisting into the
graph database took \textit{1,4s} on average.



\subsection{Identifying Weaknesses using the Testbed}
\label{sec:example}

\subsubsection{Data exposure} Developers may assume that a storage they write sensitive data to is only accessible to
authorized users. Cloud architects may assume that a certain public storage will only contain uncritical data without
knowing about all applications that actually write to that storage. Listing~\ref{lst:storage-calls} shows a query to
identify such cloud resources.


\begin{lstlisting}[breaklines=true,language=sql,label=lst:storage-calls,caption=Cypher query to identify all storage
   requests \textit{rq} from any cloud resources \textit{r1} to \textit{r2} which
   have \textit{NoAuthentication} and are publicly accessible.]
MATCH p=(r1:CloudResource)<-[:SOURCE]-
 (rq:ObjectStorageRequest)-[:TO]->(r2:Storage)--
 (:HttpEndpoint)-[:AUTHENTICITY]-(:NoAuthentication)
WHERE rq.type = "append" RETURN p
\end{lstlisting}

When executed in the testbed, this returns a path from \textit{kubernetes-logs} to \textit{am-containerlog} which we
intentionally left unprotected ($DataExposure_1$, $DataExposure_2$). A more granular flow can be detected using the
query in Listing~\ref{lst:expressions-storage}, which identifies the individual expressions, such as variables that
were written to the storage. In the case of the deployed application, this includes accessing the field
\textit{requests.values} in our modified \textit{login()} function, representing the contents of the HTTP POST request.
Thus, we can discover that the credentials passed in the login function are accidentally written to an unprotected
location, accessible from the Internet ($DangerousLog_1$, $DataExposure_3$). 

\begin{lstlisting}[breaklines=true,language=sql,label=lst:expressions-storage,caption={Cypher query to identify dataflows from 
  any source code expression to any storage resource \textit{s}, which are publicly accessible.}]
MATCH p=(e:Expression)-[:DFG*]->(s:ObjectStorage)--
  ()-[:AUTHENTICITY]-(:NoAuthentication) RETURN p
\end{lstlisting}

\subsubsection{Encryption Configuration} \label{sec:enc} As a major requirement in most security and privacy
regulations, such as the General Data Protection Regulation (GDPR), the detection of proper encryption of cloud
resources is paramount in any major cloud service deployment. It can, however, be a tedious task to keep track of the
at-rest-encryption configuration of various cloud resources. As seen in Listing~\ref{lst:tls-encryption}, the CloudPG
can easily be used to find all nodes in our testbed that have an HTTP endpoint and are missing a suitable TLS
configuration, such as $Confidentiality_1$. Similar queries can be used to check for the at-rest encryption
configuration of storage objects.
\begin{lstlisting}[breaklines=true,language=sql,label=lst:tls-encryption,caption=Cypher query to identify all Cloud resources which could offer transport encryption but have it disabled or improperly configured]
MATCH p=(n:Node)--(h:HttpEndpoint)--
  (te:TransportEncryption) WHERE te.enabled = 
  false OR te.tlsVersion <> "TLS1_2" RETURN p
\end{lstlisting}

\subsubsection{Data Flow Restrictions} \label{sec:dataflow-gdpr} Certain scenarios and regulations might impose
restrictions on data flows in a Cloud ecosystem. For example, it may be the case that due to GDPR regulations data must 
not leave the European Union. The CloudPG can be used to easily find
problematic data flows between Cloud resources, as Listing~\ref{lst:resource-flow} shows.

\begin{lstlisting}[breaklines=true,language=sql,label=lst:resource-flow,caption={Cypher query to identify data flows (\textit{DFG} edges)
  between any cloud resources in different geographic regions (\textit{l1}, \textit{l2})}]
MATCH p=(l1:GeoLocation)--(:CloudResource)-
  [:DFG]-(:CloudResource)--(l2:GeoLocation)
WHERE l1 <> l2 RETURN p
\end{lstlisting}

This query returns a path between our container images, which are stored in a GitHub Container Registry
hosted in the US, and the deployed application based on the image, which is hosted in Europe ($Dataflow_2$).

A more complex example can be found in Listing~\ref{lst:app-flow}: in this case, we are looking for data flows that
originate out of an application that is deployed in location \textit{l1}. An application itself is not a cloud resource
and does not have a location, unless it is deployed on a \textit{Compute} resource (denoted by \textit{RUNS\_ON}).
Therefore, traditional static analysis tools cannot determine such data flows. Each application has a list of
functionalities, e.g. HTTP requests to other resources, which we further want to filter to only those that are connected
to applications that offer a matching HTTP endpoint and are located in \textit{l2}. Finally, we want to only select
those nodes that differ in location (\textit{l1 <> l2}). In the testbed, this yields the problematic flow from the
US-based AWS EC2 VM to our micro-services deployed in Europe ($Dataflow_1$).

\begin{lstlisting}[breaklines=true,language=sql,label=lst:app-flow,caption=Cypher query to track data flows between an application and a Cloud resource in different geographic regions]
MATCH p=(l1:GeoLocation)-[]-(:Compute)-[:RUNS_ON]-
  (:Application)-[]-(r:HttpRequest)-[:TO]-
  (e:HttpEndpoint)-[*2]-(:Application)-[:RUNS_ON]-
  (:Compute)-[]-(l2:GeoLocation)
WHERE l1 <> l2 RETURN p 
\end{lstlisting}

\subsection{Fulfillment of the Design Goals}


\begin{table}
  \caption{Comparison of the Weaknesses identified by the CloudPG and other classes of tools}
  \centering
  \label{tab:comparison}
  \begin{tabular}{|l|c|c|c|}
    \hline
    \textbf{Weakness ID} & \textbf{CloudPG} & \textbf{SAST} & \textbf{Infrastructure Monitoring}\\
    \hline
    $DataExposure_1$ & x & - & x\\
    \hline
    $Confidentiality_1$ & x & - & x\\
    \hline
    $DataFlow_1$ & x & - & -\\
    \hline
    $DataFlow_2$ & x & - & x\\
    \hline
    $DangerousLog_1$ & x & x & - \\
    \hline
    $DataExposure_2$ & x & - & x \\
    \hline
    $DataExposure_3$ & x & - & -\\
    \hline
  \end{tabular}
\end{table}

\subsubsection{Analyze cloud-hosted code independently of the resource type or cloud provider (\textbf{DG1})}
\label{sec:lim-ontology}

Using the CloudPG, weaknesses and data-flows can be identified regardless of the underlying resource type or cloud
provider. We achieve this by using an ontology-based approach focused on the resources' functionalities and security features. 
With regards to the instantiation of the ontology, we only focus on a subset of services, mainly related to popular ones
such as VM computing, storage and some network devices. Within this scope, the abstraction works well across similar
cloud providers, such as AWS and Azure and is easily extendible in the future.

\subsubsection{Bridge the gap between static code analysis and runtime assessment (\textbf{DG2})}

The evaluation testbed shows that the CloudPG is effective in identifying security threats that result from a
misalignment of code and runtime properties which would not be possible by applying code analysis or infrastructure
monitoring alone. For example, a regular SAST tool would have only detected that a potentially dangerous log operation
is executed in weakness $DangerousLog_1$, but only the additional environment context of the CloudPG can detect that
this leads to an exposure of login credentials ($DataExposure_3$). Table~\ref{tab:comparison} shows a comparison of the
CloudPG and other classes of tools with regards to the possible detection of the weaknesses in the testbed.

\subsubsection{Track end-to-end data flows across cloud resources (\textbf{DG3})}

Section~\ref{sec:dataflow-gdpr} demonstrates the CloudPG's effectiveness in identifying problematic end-to-end data
flows in a heterogeneous cloud service even across different cloud vendors. To the best of the
authors' knowledge, no other approach and implementation has been proposed before.


\subsubsection{Provide an efficiently searchable representation of the results (\textbf{DG4})}

The prototype implementation used the Cypher language to interact with the persisted CloudPG. It is a feature rich query
language for graph databases and all weaknesses introduced in the testbed were found using complex queries. All queries
returned their results in less than 5 ms, making it suitable for a CI/CD environment or even for use during the development 
process in an IDE.



%% file: sections/related.tex

Various works have proposed systematizations for cloud resources and cloud security threats. Joshi et al.
\cite{joshi20integrated} propose a knowledge graph that models various compliance requirements and also maps them to
security controls and threats. Others have proposed systematizations of cloud infrastructure offerings, e.g. Sikeridis
et al. \cite{sikeridis17comparative}. Hendre and Joshi \cite{hendre2015semantic} have proposed an ontology of security
controls, threats, and security-related standards, like ISO standards. Iqbal et al. \cite{iqbal2016cloud} propose a
taxonomy for attacks on cloud systems and differentiate between attacks on the different service models. Contrary to the
approaches mentioned above, we aim to specifically model those security features that can be configured on the
management plane and which we want to represent in the graph representation afterwards.


There are various approaches and tools to monitor public cloud resources, e.g., GmonE \cite{montes2013gmone}, Clouditor
\cite{clouditor-wp}, specific approaches for OpenStack \cite{AnisettiADGV15}, and many others \cite{ward2014observing}.
Yet, while they automatically discover existing resources and their configurations, the expected secure configurations
usually need to be specified manually since they do not build upon an abstract taxonomy of cloud resources and their
security features. In the proposed approach, we enable a query-based evaluation of a system's security posture using the
graph structure the cloud property graph creates.

Automated threat analysis of cloud systems has been tackled by An et al. \cite{an2019cloudsafe} who propose the
CloudSafe tool. It combines an automatically generated reachability graph of VMs with known vulnerabilities to identify
security threats. Other approaches similarly assess risks using known vulnerabilities, for instance Kamongi et al.
\cite{kamongi2014nemesis}. In summary, existing approaches have not addressed the problem of connecting configuration
monitoring with static code analysis to identify security problems resulting from misconfigurations and inconsistencies
between application-level and configuration-level assumptions.


%% file: sections/conclusions.tex

In this paper, we have presented the Cloud Property Graph, an extension of a code property graph, based on a
comprehensive ontology of cloud resources and their generalized functionalities and security features (\textbf{DG1}). It
bridges the gap between static code analysis and runtime assessment of cloud services (\textbf{DG2}) by providing
additional context from runtime configuration information. Thus, our CloudPG-based analysis framework can be used by
security experts to quickly identify weaknesses in their cloud deployments based on generalized feature configurations,
rather then specifics of an individual cloud vendor. This includes tracking of data flows across applications and cloud
resources (\textbf{DG3}). We have shown in an evaluation testbed how these design goals are applicable to a target cloud
service and have proposed example queries (\textbf{DG4}) for several common weaknesses.

For future work, we are planning various extensions of the CloudPG and the ontology. 
We want to extend the instantiated ontology in a community approach, possibly with more frameworks from other languages.
Lastly, we aim at building queries for more classes of weaknesses and explore the option to exploit the CloudPG for
assessing privacy risks in the Cloud.